\documentstyle[10pt]{article}

\begin{document}

\title{A Proposed Experiment Showing that Classical Fields Can Violate
Bell's Inequalities}

\author{{\bf Patrick Suppes}\thanks{E-mail:
suppes@ockham.stanford.edu. To whom correspondence should be
addressed.} \and {\bf J. Acacio de Barros}\thanks{Permanent Address:
Physics Department, Federal University at Juiz de Fora, 36036-330 Juiz
de Fora, MG Brazil. E-mail: acacio@fisica.ufjf.br} \and {\bf Adonai
S. Sant'Anna}\thanks{Permanent Address: Mathematics Department,
Federal University at Paran\'a, C.P. 19081, 81530-900, Curitiba, PR,
Brazil. E-mail: adonai@gauss.mat.ufpr.br} \\ {\it Ventura Hall,
Stanford University,} \\ {\it Stanford, California 94305-4115}}

\date{\today}

\maketitle

\newcounter{cms}

\setlength{\unitlength}{1mm}

\begin{abstract}

We show one can use classical fields to modify a quantum optics
experiment so that Bell's inequalities will be violated. This happens
with continuous random variables that are local, but we need to use
the correlation matrix to prove there can be no joint probability
distribution of the observables. 

\end{abstract}

\vspace{1em}

\section{Introduction}

The issue of the existence of hidden variables for quantum mechanics is
almost as old as quantum mechanics itself. However, in 1963 J. S. Bell
showed \cite{Bell-65,Bell-66} that if one makes some ``reasonable''
assumptions about the hidden variables, like locality and statistical
independence of distant measurements, the correlations for the outcome
of measurements for an EPR-like experiment have to satisfy a set of
inequalities. In 1982 Alain Aspect and coworkers showed that quantum
mechanics violated Bell's inequalities, drawing the conclusion
that one cannot have a local realistic theory that would replace
quantum mechanics \cite{Aspect1,Aspect2}.

Because Bell's assumptions were considered equivalent to the existence
of an underlying physical reality, it is often said that any classical
system satisfies Bell's inequalities. In this paper we will show that
classical fields {\em do not} satisfy Bell's inequalities, hence
classical fields, e.g. electromagnetic fields, are not Bell-type
hidden variables. We do this by showing that a simple experimental
setup, suggested by Tan {\em et al.} \cite{Tan,Walls}, can be
reinterpreted for classical electromagnetic fields. For this
reinterpretation we derive from the classical field properties a
violation of Bell's inequalities\cite{Bell-65,Bell-66,Suppes-76},
with, at the same time, locality being preserved in a sense to be made
precise.

\section{Experimental Setup}

The experimental scheme uses two classical coherent sources
$\alpha_1(\theta_1)$, with phase $\theta_1$, and $\alpha_2(\theta_2)$,
with phase $\theta_2$, and a third source to be studied, $u(\theta)$,
with unknown phase.  The experimental configuration has two homodyne
detections, $(D_1,D_2)$ being one and $(D_3,D_4)$ the other, such that
the measurements are sensitive to phase changes in $u(\theta)$.  The
geometry of the setup is shown in FIG.\ 1.  
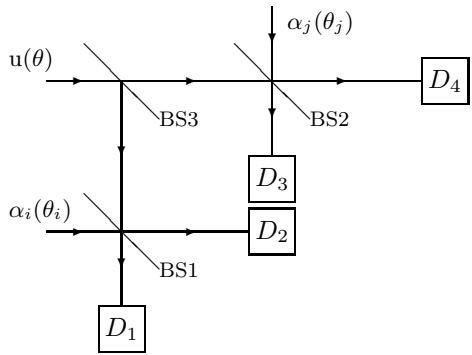
\begin{figure}[h]
\begin{picture}(100,50)
\put(40,40){\vector(1,0){5}}
\put(45,40){\vector(1,0){15}}
\put(60,40){\vector(1,0){20}}
\put(80,40){\line(1,0){10}}
\put(90,37){\framebox(6,6){$D_{4}$}}
\put(40,20){\vector(1,0){5}}
\put(45,20){\vector(1,0){15}}
\put(60,20){\line(1,0){7}}
\put(67,17){\framebox(6,6){$D_{2}$}}
\put(50,40){\vector(0,-1){10}}
\put(50,30){\vector(0,-1){15}}
\put(50,15){\line(0,-1){5}}
\put(47,4){\framebox(6,6){$D_{1}$}}
\put(70,50){\vector(0,-1){5}}
\put(70,45){\vector(0,-1){10}}
\put(70,35){\line(0,-1){5}}
\put(67,24){\framebox(6,6){$D_{3}$}}
\put(35,42){\small u($\theta$)}
\put(35,22){\small $\alpha_{i}(\theta_{i})$}
\put(72,47){\small $\alpha_{j}(\theta_{j})$}
\put(45,25){\line(1,-1){10}}
\put(45,45){\line(1,-1){10}}
\put(65,45){\line(1,-1){10}}
\put(55,34){\footnotesize BS3}
\put(75,34){\footnotesize BS2}
\put(55,14){\footnotesize BS1}
\end{picture}
\caption{Proposed experimental configuration.}
\end{figure}
In FIG.\ 1 BS1, BS2 and BS3 are beam splitter mirrors
that will reflect 50\% of the incident electromagnetic field and let
50\% of it pass. When the electromagnetic field is reflected, the
mirrors add a phase of $\pi/2$ to the field, while no phase is added
when the field passes through BS1, BS2 or BS3. We will look for
correlations between the pairs of detectors $(D_1,D_2)$ and
$(D_3,D_4)$.

\section{Correlation Functions}

In this section we compute the correlation functions that violate the inequalities. We first define the random 
variables in terms of which we derive the
Bell-type correlations. On this matter we shall be as explicit as
possible. Associated to the source $u(\theta)$ at $D_1$ is the random
variable $U_{1}(t)$, whose value at $t$ is just the value of the
classical field at $D_1$, namely,
\begin{equation}
U_{1}(t) = \frac{1}{4}\beta\cos (\omega t + \theta + \pi/2),\label{U1}
\end{equation}
where $\beta$ is the amplitude of the field at the source, $\theta$ is
the unknown phase and $\pi/2$ is a phase gained when $u$ is reflected
at BS3.

Probability enters by using the time average to compute the
expectation of $U_{1}(t)^2$
\begin{equation}
U_1^2 = \langle U_{1}(t)^2 \rangle = \langle [\frac{1}{4}\beta\cos
(\omega t +
\theta + \pi/2)]^2 \rangle ,
\end{equation}
which is just the standard intensity, but here we treat it
probabilistically. In a similar fashion, associated to the source
$\alpha_1(\theta_1)$ at $D_1$ is the random variable $A_1(t)$,
\begin{equation}
A_1(t) = \frac{1}{2}\alpha\cos (\omega t + \theta_1 + \pi/2)
\end{equation}
 and thus
\begin{equation}
A_1^2 = \langle A_1(t)^2 \rangle = \langle [\frac{1}{2}\alpha\cos
(\omega t +
\theta_1 + \pi/2)]^2 \rangle .
\end{equation}

At $D_1$, the total field is the random variable $F_1(t) =
U_{1}(t)+A_1(t)$. So, the intensity of the total field at $D_1$ is
just the second moment of $F_1(t)$, i.e.,
\begin{eqnarray}
I_1(\theta) & = & F_1^2 = \langle F_1(t)^2 \rangle = \langle
(U_{1}(t)+A_1(t))^2
\rangle \nonumber \\
	    & = & \langle U_{1}(t)^2 \rangle + 2 \langle
(U_{1}(t)A_1(t)) \rangle +\langle A_1(t)^2 \rangle,
\end{eqnarray}
where we used $\theta$ as an argument for $I_1$ to make it explicit
that it depends on $\theta$. We can see that the cross moment in the
expression above is the classical interference term.

We can compute $I_1$ directly from the expression for $U_{1}(t)$ and
$A_1(t)$ in the following way
\begin{eqnarray}
I_1(\theta) & = & \lim_{T\rightarrow\infty}  \frac{1}{T}
\int_0^T [\frac{1}{2}\alpha\cos (\omega t + \theta_1 + \pi/2) +
\nonumber \\ & & \frac{1}{4}\beta\cos (\omega t + \theta + \pi/2)]^2dt ,
\end{eqnarray}
 which is
\begin{equation}
\label{seven}
I_1(\theta)= \frac{1}{32}\beta^2 + \frac{1}{8}\alpha\beta\cos
(\theta-\theta_1)
+ \frac{1}{8}\alpha^2.
\end{equation}

In similar fashion, we can compute for the other three detectors,
\begin{equation}
\label{eight}
I_2(\theta) = \frac{1}{32}\beta^2 - \frac{1}{8}\alpha\beta\cos
(\theta-\theta_1)
+ \frac{1}{8}\alpha^2,
\end{equation}
\begin{equation}
I_3(\theta) = \frac{1}{32}\beta^2 - \frac{1}{8}\alpha\beta\sin
(\theta-\theta_2)
+ \frac{1}{8}\alpha^2,
\end{equation}
and
\begin{equation}
\label{ten}
I_4(\theta) = \frac{1}{32}\beta^2 + \frac{1}{8}\alpha\beta\sin
(\theta-\theta_2)
+ \frac{1}{8}\alpha^2.
\end{equation}

The intensities obtained are conditional on $\theta$. To obtain the
unconditional intensities we assume a uniform distribution for
$\theta$ and integrate the expressions for all possible values of
$\theta$. Not only is $\theta$ unknown, but the phase would vary
randomly for repeated runs of the experiment. If $\theta$ were a
coherent source with fixed $\theta$, Bell's inequalities would not be
violated \cite{Tan90}. 

The unconditional intensities $I_1$, $I_2$, $I_3$, and $I_4$ for the
detectors $D_1$, $D_2$, $D_3$, and $D_4$ are
\begin{equation}
\label{pD1wotheta}
I_1 = \frac{1}{32}\beta^2 + \frac{1}{8}\alpha^2,
\end{equation}
\begin{equation}
\label{pD2wotheta}
I_2 = \frac{1}{32}\beta^2 + \frac{1}{8}\alpha^2,
\end{equation}
\begin{equation}
\label{pD3wotheta}
I_3 = \frac{1}{32}\beta^2 + \frac{1}{8}\alpha^2,
\end{equation}
\begin{equation}
\label{pD4wotheta}
I_4 = \frac{1}{32}\beta^2 + \frac{1}{8}\alpha^2.
\end{equation}
We can see from (\ref{pD1wotheta})--(\ref{pD4wotheta}) that the
intensities are the same for all detectors, and are similar to those
given by Walls and Milburn \cite{Walls} in the case of a classical
source.

We now start computing the covariance between intensities in the
homodyne detectors. The covariance we are interested in is between
$(I_1-I_2)$ and $(I_3-I_4)$. 
\begin{eqnarray}
\label{15}
\mbox{Cov}(I_1-I_2,I_3-I_4)  =  \frac{1}{2\pi}\int_{0}^{2\pi}[(I_1(\theta) - I_2(\theta))\times (I_3(\theta) - I_4(\theta))] d\theta  \nonumber  \\
 -\ \frac{1}{2\pi}\int_{0}^{2\pi} (I_1(\theta)-I_2(\theta)) d\theta \times 
 \ \frac{1}{2\pi}\int_{0}^{2\pi} (I_3(\theta)-I_4(\theta)) d\theta. 
\end{eqnarray}
It is straightforward to show from (\ref{seven})--(\ref{ten}) and
(\ref{15}) that
\begin{equation}
\mbox{ Cov}(I_1-I_2,I_3-I_4) =
-\frac{1}{32}\beta^2\alpha^2\sin(\theta_1-\theta_2).
\end{equation}

In order to compute the correlation we have to know the variance of
the random variables $(I_1-I_2)$ and $(I_3-I_4)$, which are defined as
\begin{equation}
\mbox{ Var}(I_1-I_2) = 
\frac{1}{2\pi}\int_{0}^{2\pi}(I_1(\theta)-I_2(\theta))^2 d\theta -
 \left[ \frac{1}{2\pi}\int_{0}^{2\pi}(I_1(\theta)-I_2(\theta))
d\theta \right]^2 = \frac{1}{32}\beta^2\alpha^2
\end{equation}
and 
\begin{equation}
\mbox{ Var}(I_3-I_4)  =
\frac{1}{2\pi}\int_{0}^{2\pi}(I_3(\theta)-I_4(\theta))^2 d\theta -
 \left[\frac{1}{2\pi}\int_{0}^{2\pi}(I_3(\theta)-I_4(\theta)) d\theta
\right]^2 = \frac{1}{32}\beta^2\alpha^2.
\end{equation}
Finally, we are in a position to compute the correlation between the
two random variables $(I_1-I_2)$ and $(I_3-I_4)$. This is done in the
standard way, by just dividing the covariance by the squareroot of the
variances:
\begin{equation}
\rho (I_1-I_2,I_3-I_4) = \frac{\mbox{ Cov}(I_1-I_2,I_3-I_4)
}{\sqrt{\mbox{ Var}(I_1-I_2) \mbox{ Var}(I_3-I_4) }},
\end{equation}
and we have the following expression for the correlation
\begin{equation}
\rho (I_1-I_2,I_3-I_4) = -\sin (\theta_1 - \theta_2),
\end{equation}
which we may rewrite as 
\begin{equation}
\rho (\theta_1,\theta_2) 
= -\sin (\theta_1 - \theta_2).\label{correlationrho}
\end{equation}

\section{Violation of Bell's Inequalities}

We are now in a position to show that we can violate Bell's inequalities. 
We may now choose angles $\theta_1$, $\theta_2$, $\theta_{1}'$, and
$\theta_{2}'$ such that we obtain at once, for the four correlations
$\rho (\theta_1,\theta_2)$, $\rho (\theta_{1},\theta_{2}')$, $\rho
(\theta_{1}',\theta_2)$ and $\rho (\theta_{1}',\theta_{2}')$ a
violation of Bell's inequalities in the form due to Clauser, Horne and
Shimony
\cite{Clauser-78}, by choosing the four angles such that
\begin{equation}
\theta_{1} - \theta_{2} = \theta_{1}' - \theta_{2}' = 60^{o} ,
\end{equation} 
\begin{equation}
\theta_1 - \theta_{2}' = 30^{o} ,
\end{equation} 
\begin{equation}
\theta_{1}' - \theta_2 = 90^{o} .
\end{equation} 
In particular, 
\begin{eqnarray}
\rho (\theta_1,\theta_2) - \rho (\theta_1,\theta_2') + \rho
(\theta_1',\theta_2) + \rho (\theta_1',\theta_2') & = & \nonumber \\
-\frac{\sqrt{3}}{2} + \frac{1}{2} - 1 - \frac{\sqrt{3}}{2} & < & -2.
\end{eqnarray}

	However, in the case of continuous random variables, which is
what we have in the present context for intensity, or differences of
intensity, failure to satisfy Bell's inequalities in the Clauser,
Horne and Shimony form does not imply that there can be no joint
distribution of the four random variables compatible with the four
given correlations. In fact, it is easy to show that for selected
values of the two missing correlations, there does, for this example,
exist a joint probability of the four random variables compatible with
the four given correlations. What is required in the general case, as
opposed to that of discrete $\pm 1$-values, to test for the existence
of a joint distribution, when means and correlations are given, is
that the eigenvalues of the correlation matrix are all
nonnegative. This extends the earlier result of \cite{Fine} for
discrete $\pm 1$ values. Because of the freedom to select arbitrarily
the two missing correlations in the Clauser, Horne and Shimony form of
Bell's inequalities, we have not been able to construct an example
using $-\sin(\theta_{i}-\theta_{j})$ for the correlations that has at
least one negative eigenvalue for all possible values of the missing
correlations.

	Another possibility is obvious. Bell's original paper
\cite{Bell-65} used three rather than four random variables, and, put
in terms of this letter, he showed that the correlations for the three
angle differences derived from $\theta_{1}$, $\theta_{2}$, and
$\theta_{3}$ violated the following inequality, necessary for the
existence of a joint distribution of three discrete random variables
with values $\pm 1$: \begin{equation} \rho(\theta_{1},\theta_{2})+
\rho(\theta_{1},\theta_{3})+ \rho(\theta_{2},\theta_{3}) \geq
-1.\label{inequalitywiththree} \end{equation} To violate
(\ref{inequalitywiththree}) we choose three angles $\theta_{1}$,
$\theta_{2}$ and $\theta_{3}$, with \begin{equation} \theta_{1} =
0,\label{theta1} \end{equation} \begin{equation} \theta_{2} = 45^{o},
\end{equation} \begin{equation} \theta_{3} = 90^{o},\label{theta3}
\end{equation} and, using equation (\ref{correlationrho}), the
correlation matrix is \begin{equation} \left(\begin{array}{ccc} 1 &
-\sqrt{2}/2 & -1\\ -\sqrt{2}/2 & 1 & -\sqrt{2}/2\\ -1 & -\sqrt{2}/2 &
1\\ \end{array}\right).\label{matrix} \end{equation}

	It is a direct computation to show this matrix has both
positive and negative eigenvalues, so that it is not nonnegative
definite. The eigenvalues are $(\sqrt{5}+1)/2$, $(-\sqrt{5}+1)/2$ and
$2$. Therefore, there can be no joint probability distribution for the
three random variables compatible with the correlations given in
(\ref{matrix}). From results in \cite{SuppesZanotti81} and
\cite{Holland} there can then be no hidden variable that factors out
the correlations conditionally, i.e., there can be no $\lambda$ such
that for the three random variables $X(\theta_{1})$, $Y(\theta_{2})$
and $Z(\theta_{3})$, we have \begin{equation} E(XYZ|\lambda) =
E(X|\lambda)E(Y|\lambda)E(Z|\lambda), \end{equation} since there is no
joint probability distribution of $X$, $Y$ and $Z$ compatible with the
given correlations. In particular, $\theta = \lambda$ cannot serve as
a Bell-type hidden variable for a classical field described by
(\ref{U1}).

	Finally, we note that even though (\ref{inequalitywiththree})
was violated by the angle values in (\ref{theta1})-(\ref{theta3}),
this inequality is not a satisfactory general test for existence of a
joint distribution, as the following shows. Let $\theta_{1} = 0$,
$\theta_{2} = 30^{o}$ and $\theta_{3} = 45^{o}$. Then it is easy to
check that inequality (\ref{inequalitywiththree}) is violated, but the
eigenvalues of the correlation matrix are all nonnegative, and so a
joint distribution exists.

\section{Measurement and Photon Counts} 

Because classical field theory is a
deterministic theory, our introduction of expectations and
probabilities might be questioned. Our response is that the strength
of a classical field at a space-time point cannot be measured, as was
emphasized long ago by Bohr and Rosenfeld in a famous paper in 1933
\cite{BohrRosenfeld}. As they pointed out, classical field strength
cannot be represented by true point functions, but by average values
over space-time regions. This is exactly what we have done in
introducing random variables and their expectations. The casual reader
might claim that we should do  an analysis of coincidence counts with
photocounters. This makes no sense in the case of classical
fields, where the number of photons arriving at the same time at each
detector is incredibly large. What makes sense is not discrete but
continuous measurement of intensity. 

Despite that, we are going to use the previous result to model
discrete photon counts in such a way that they violate Bell's
inequalities.  For this, we define two new discrete random variables
$X=\pm 1$ and $Y=\pm 1$. These random variables correspond to nearly
simultaneous correlated counts at the detectors, and are defined in
the following way.
\begin{equation}
X = \left\{  \begin{array}{ll}
	    +1 & \mbox{if detector $D_1$ triggers a count} \\
	    -1 & \mbox{if detector $D_2$ triggers a count} 
	    \end{array}
     \right.
\end{equation}
\begin{equation}
Y = \left\{  \begin{array}{ll}
	    +1 & \mbox{if detector $D_3$ triggers a count} \\
	    -1 & \mbox{if detector $D_4$ triggers a count.} 
	    \end{array}
     \right.  \end{equation} To compute the expectation of $X$ and $Y$
we use the stationarity of the process and do the following. First,
let us note that \begin{equation} I_1-I_2 = N_X \cdot P(X=1) - N_X
\cdot P(X=-1), \end{equation} where $N_X$ is the expected total number
of photon counts at $D_1$ and $D_2$ and $P(X=\pm 1)$ is the
probability that the random variable $X$ has values $\pm 1$. The same
relation holds for \begin{equation} I_3-I_4 = N_Y \cdot P(Y=1) - N_Y
\cdot P(Y=-1).  \end{equation} To simplify we put as a symmetry
condition that $N_X=N_Y=N$, i.e., the expected number of photon counts
at each homodyne detector is the same. But we know that
\begin{equation} I_1 + I_2 = N \cdot P(X=1) + N \cdot P(X=-1) = N,
\end{equation} and 
\begin{equation} I_3 + I_4 = N \cdot P(X=1) + N
\cdot P(X=-1) = N.  
\end{equation} 

Then we can conclude from equations (\ref{seven})---(\ref{ten}),
assuming maximum visibility, that
\begin{equation} E_d(X|\theta) = \frac{I_1 - I_2}{I_1 + I_2} =
\cos(\theta-\theta_i), \end{equation} \begin{equation} E_d(Y|\theta) =
\frac{I_3 - I_4}{I_3 + I_4} = \sin(\theta-\theta_j), \end{equation}
where $E_d$ represents the expected value of the counting random
variable. It is clear that if $\theta$ is uniformly distributed we
have at once: \begin{equation} E(X) = E_\theta(E_d(X|\theta)) = 0,
\label{expectX} \end{equation} \begin{equation} E(Y) =
E_\theta(E_d(X|\theta)) = 0.  \label{expectY} \end{equation} We can
now compute Cov$(X,Y)$. Note that \begin{eqnarray} \mbox{Cov}(X,Y) & =
& E(XY) - E(X) E(Y) \nonumber \\
		& = & E_\theta(E_d(XY|\theta)) - 
		      E_\theta(E_d(X|\theta))
		      E_\theta(E_d(Y|\theta))
\end{eqnarray}
and so
\begin{eqnarray}
\mbox{Cov}(X,Y) & = & \frac{1}{2\pi} \int^{2\pi}_{0} E_d(XY|\theta)
 d\theta \nonumber \\ & & - \frac{1}{2\pi} \int^{2\pi}_{0}
 E_d(X|\theta) d\theta \times \frac{1}{2\pi} \int^{2\pi}_{0}
 E_d(Y|\theta)d\theta.
\end{eqnarray}
In order to compute the covariance, we also use the conditional
independence of $X$ and $Y$ given $\theta$, which is our locality
condition:
\begin{equation}
E_d(XY|\theta) = E_d(X|\theta) E_d(Y|\theta),
\end{equation}
because given $\theta$, the expectation of $X$ depends only on
$\theta_i$, and of $Y$ only on $\theta_j$. Then, it is easy to see
that
\begin{equation}
\rho (X,Y) = \mbox{Cov}(X,Y) = - \sin(\theta_i-\theta_j).
\label{finalcorrelation}
\end{equation}
The correlation equals the covariance, since $X$ and $Y$ are discrete
$\pm 1$ random variables with zero mean, as shown in (\ref{expectX})
and (\ref{expectY}), and so $\mbox{Var}(X) = \mbox{Var}(Y)  = 1.$ It
follows at once from (\ref{finalcorrelation}) that for a given set of
$\theta_i$'s and $\theta_j$'s  Bell's inequalities are violated.

\section{Locality} 

Much of the discussion involving Bell's theorem is
connected to locality. For that reason, we will prove in this section
that the scheme presented in this paper is local in one precise sense.
We follow \cite{Suppes-76}. Locality requires the following:
\begin{equation}
\label{locality}
E(X|\theta_{1},\theta_{2},\theta) = E(X|\theta_{1},\theta).
\end{equation}
It is obvious that (\ref{locality}) follows immediately from subtracting 
(\ref{eight}) from (\ref{seven}), and observing the result does not depend 
on $\theta_{2}$, and similarly for the other cases of random variables $Y$ 
and $Z$. Equation (\ref{locality}) says
simply that whatever is the result of the measurement at one homodyne
detector, it must depend only on $\theta$, the hidden variable, and
the phase associated to this particular detector, and cannot be
influenced by the phase at the other detector.

\section{Proposed experiment.}  

The experiment proposed in \cite{Tan} supposes a single photon source
that is split into the two homodyne detectors. Tan {\em et al.} also
analyze the classical case and get no violation of Bell's
inequalities.  However, they assume a weak coherent source with
randomized phase as the classical analogue of their single photon
source. This would be equivalent to having a classical thermal source,
where coherence would not be a strong feature. In our experiment we
suppose that this source is not only classical, i.e., with high
intensity, but also that it is coherent with the phase unobservable
and varying randomly on repeated runs. The different source used here,
as opposed to that used in \cite{Tan} implies that the expectations
given by (7)--(14) are computed in a different way than in
\cite{Tan}. Here we first integrate with respect to $t$ and them
integrate with respect to $\theta$. It is easy to supply a source
that would fit our requirements. This would be, for example, a radio
source, a microwave, or a laser source, all with unstabilized
phases. To realize this experiment, one must also use two additional
coherent sources with stable known phases and with the same frequency
as the nonstabilized source. If a data table is then built that keeps
track of all the measured values on the detectors, we can compute the
correlations and see a violation of Bell's inequalities, or do the
stronger test using the 3-variable version and the matrix
(\ref{matrix}).

\section{Final Remarks}

There are several remarks that we must add in order to clarify some
points. 

First, when using classical fields the number of photons is
overwhelmingly large. For that reason, we would not need to compute any
photon count correlation. What we measure is intensity. On the other
hand, Bell's inequalities are not enough to show that we do not have a
joint probability distribution for classical fields, because they
assume a continuous range of values. That is why we computed the
correlation matrix, showing that for this case a joint probability
distribution does indeed not exist. 

Another point is that intensity of classical fields does not satisfy
the basic assumption made by Bell, because it can take an infinite
range of values; Bell considered spin measurements that can take only
two possible values. To show that this does not present any
constraint, in Section V we did an analysis of photon counts, which
can only take, as in Bell's assumptions, two discrete values.

Finally, the last point. One can argue that if classical fields
violate Bell's inequalities, then, since they are classical, Bell's
theorem must be wrong, and we must show why it is wrong. We did not
show that Bell's theorem is wrong. We just showed that a classical
field is not a Bell-type hidden-variable. What is wrong is the
preconception that anything classical must satisfy Bell's hypothesis.

\vspace{1em}

\paragraph{Acknowledgments.} J.~A.~B. acknowledges support from the
Department of Fields and Particles (DCP) and the Laboratory for HEP
and Cosmology (Lafex) of the Brazilian Centre for Physical Research
(CBPF). A.\ S.\ S. wishes to thank CNPq (Brazilian Government Support
Agency) for partial financial support.

\end{document}